\documentclass[12pt]{article}
\usepackage{amssymb}
\usepackage{amsfonts}
\usepackage{amsmath,amsthm}
\usepackage{graphics,epsfig}
\usepackage{subfigure}
\usepackage{graphicx,epsfig, color }
\oddsidemargin 10mm \numberwithin{equation}{section}
\def\be{\begin{equation}}
\def\ee{\end{equation}}
\begin{document}
\begin{center}
{{\bf {Particles creation from JNW quantum perturbed black holes
by  minimally coupled Klein Gordon scalar free fields }} \vskip
0.5 cm { Hossein Ghaffarnejad \footnote{E-mail address:
hghafarnejad@semnan.ac.ir} and~~Hamid Reza Faghani
\footnote{E-mail
address: h.r.faghani@semnan.ac.ir}}} \\
\vskip 0.1 cm {\textit{Faculty of Physics, Semnan University, P.C.
35131-19111, Semnan, Iran} } \vskip 0.1 cm
\end{center}
\begin{abstract}
In this work, we choose a minimal coupling interaction between
massive Klein Gordon (KG) quantum scalar free fields and
Janis-Newman-Winicour (JNW) spherically symmetric static black
hole, to produce its Hawking temperature and luminosity. This is
done by calculating asymptotic wave solutions at near and far from
the black hole horizon. They are orthogonal mode solutions of
local Hilbert spaces. By using these mode solutions, we calculated
Bogolubov coefficients and then, we investigated number density
matrix of created particles. Mathematical calculations show that
this is not exactly similar to the Planck`s black body radiation
energy density distribution but, it is "gray" body radiation
distribution depended to the emitted Hawking particles frequency.
Their difference is a non-vanishing absorptivity factor of
backscattered particles after to form horizon of a collapsing
body. Our motivation is determination of position of Hawking
created pairs in which, two different proposals are proposed, so
called as "fairwall" and "quantum atmosphere".
\end{abstract}
\section{Introduction}
In absence of pure quantum gravity theory \cite{Fauser}, the
quantum matter fields theories in curved space times
\cite{Bir},\cite{Par} are applicable in studying the behavior of
quantum astrophysical objects such as black holes. The
semiclassical framework of quantum gravity is a perturbative
approach for unknown quantum gravity theory and it is valid only
for scales higher than the Planckian scales
$\ell_p=1.6\times10^{-35} m,$
$t_p=5.4\times10^{-44}s$ and $T_p=1.4\times10^{32} K^{\circ}.$\\
 As an application of the semiclassical approach of quantum gravity on primordial black
 holes,
we should address most famous discovery of Stephen Hawking at 1974
which at a first time, he showed thermal radiation of the
primordial black holes \cite{Haw1}, \cite{Haw2}. This discovery
changed our perspective towards black holes, giving us a deeper
insight about the microscopic nature of gravity. This raises many
deep puzzles. Among them, are whether black holes can evaporate
completely or
must they leave relics?\\
If the former case is true, what has happened to any conserved
quantum numbers? Perhaps black holes evaporation processes and
their virtual counterparts will violate any global conservation
law. More puzzling still: what happens to quantum coherence during
this process. Where does the information go? This is so called now
as the information loss paradox \cite{Haw3}, \cite{Page},
\cite{Gidd1}. Namely, the black hole thermal radiation causes that
there would be no $S$ matrix to take an initial pure state to a
final pure state. There is no unitary evolution in black holes
quantum mechanics same as one prescribed in the ordinary quantum
mechanics. Because, the black hole would eventually disappear
completely and that the resulting state of radiation, like a
precisely thermal state, would be mixed. At the same time, within
the semi-classical framework, the current understanding of such
process still leaves open several issues.\\
If the latter case is true, then the semiclassical approach of
unknown quantum gravity is invalid at sub-Plankian scales and its
predictions valid just for scales larger than the Plankian.
Although an accurate and general answer based on pure quantum
gravity has not yet been provided, research has been done in this
direction with the approach of quantum field perturbation theory
in curved space: See for instance \cite{Ghaf1} for study of
effects of cosmological constant and back-reaction of massless
quantum scalar fields on evaporating Schwarzschild- de Sitter
black hole and see \cite{Ken}, \cite{Rali} for investigation of
effects of higher dimensions in produce of remnant stable final
state of quantum evaporating black hole. Also, \cite{Hir} and
\cite{Elis} are suitable works in which effects of expanding
universe is considered to control the Hawking radiation of a
stationary black hole such that it remains as
stable.\\
In fact, many of these problems cannot be solved within the
semiclassical quantum gravity and nowadays their answers are
investigated in string theories \cite{Barton}, \cite{Gasperini} or
its other branches such as M
theory, holography caused by AdS/CFT correspondence \cite{Gibb}.\\
Discovery of the Hawking radiation of black holes raises a new
branch of physics called as black hole thermodynamics. This caused
to have three different methods to calculate the Hawking
temperature of a quantum evaporating black hole. They are as
follows: (a) Calculation of renormalized expectation value of
stress tensor operator of created particles on the black hole
spacetime where by regarding the covariant conservation of
renormalized stress tensor, we obtain a non vanishing trace
anomaly for the renormalized expectation value of the stress
tensor operator which reads to the Hawking radiation in the black
hole space time (see for instance \cite{Ghaf2} and references
therein),(b) calculation of orthogonal mode solutions in
asymptotic regions of black hole spacetime and corresponding
Bogolubove coefficients  to obtain distribution function of number
density of created particles (see \cite{Bir}, \cite{Par} and
\cite{Haw2}).(c) calculation of surface gravity on the black hole
horizon to obtain equation of state of quantum black holes by
regarding their thermodynamics aspect and AdS/CFT correspondence
(More publications are in the literature where this method is used
for instance see
 \cite{Ghaf3} and references therein.).\\
The latter method is simplest with respect to (a) and (b) to
obtain the temperature, in which, we use Euclidean signature of
the black hole metric to calculate the surface gravity. For
instance, for a Schwarzschild black hole $g_{tt}=1-\frac{2M}{r}$
we obtain $T=\frac{1}{4\pi}\frac{d
g_{tt}}{dr}|_{r=2M}=\frac{1}{8\pi M}$ (in units $G=c=1$), but, as
we see later in this work, that this is not applicable for JNW
black hole and we must
 choose two other methods (a) or (b).\\
Our motivation in this work, is calculation of  quantum perturbed
JNW black hole \cite{JNW} temperature and determination of
position of the Hawking pairs production. At the present, two
different proposal are proposed to this position, so called as
"fairwall" (near the horizon) and "quantum atmosphere" (several
times the radius of the horizon) (see for instance \cite{Gidd} and
\cite{Dey}).\\ The JNW black hole is shown at a first time by
Janis-Newman-Winicour at 1968 \cite{JNW}, which is a spherically
symmetric static solution of the Einstein metric equation. This is
not a vacuum solution but, its source is a massless classical
scalar KG field with scalar charge $q.$ As a quantum perturbed
case of the JNW black hole, we consider an additional quantum
massive KG scalar field which is interacting minimally with the
JNW black hole.
 To calculate the Hawking
temperature of the JNW black hole, we consider just free non
self-interacting KG fields to produce particle creation. Effects
of self interaction $\lambda \Phi^4$ on rate of the particle
creation is dedicated to our subsequent work.
 The present work is organized as follows:\\ In section 2,
  we define line element of JNW black hole together with action functional of KG quantum massive scalar fields.
We calculate asymptotically mode solutions of the KG wave
equations near and far from the JNW black hole horizon.
  In section 3, we use this mode solutions to determine the Bogolubove transformations. They relate two different
  set of mode solutions to each other. We investigate distribution of number of created particles per unit frequency
  and per unit time so called as number density
   function which gives us the JNW black hole temperature too. In this section, we find related luminosity of emitted
   particles.
 In section 4, we study mass lose of the evaporating quantum JNW black hole. In section
5, we interpret diagrams of obtained theoretical results and
  compare with other related published works.
 Section 6 is dedicated to concluding remark and future
 extensions.
\section{JNW black holes and KG quantum fields }
Extension of the Israel`s singularity theorem\footnote{Among all
static, asymptotically flat, vacuum solutions of the Einstein
equations with closed simply connected equipotential surfaces,
$g=const$, Schwarzschild's solution is the only one that has a
nonsingular event horizon g=0.}, was motivation
Janis-Newman-Winicour (JNW) to propose the spherically symmetric
metric solution of a gravitational system which is coupled  with
massless scalar fields (see \cite{JNW} for more discussion). In
their model, the exterior Schwarzschild metric is obtained but,
the event horizon (the usual $r=2m$) is not only singular but, is
also a point rather than a sphere. We call this line element as
JNW metric for which different forms are presented in the
literature. For instance one can compare \cite{JNW} with
\cite{Wym} which is proven that they are related to each other by
using suitable coordinates transformation \cite{Virb}  such that
\begin{equation}\label{ds}
  ds^{2}=\bigg(1-\frac{b}{r}\bigg)^\gamma dt^2 -\bigg(1-\frac{b}{r}\bigg)^{-\gamma} dr^2
 -\bigg(1-\frac{b}{r}\bigg)^{1-\gamma} r^2 (d\theta^2 +\sin ^2 \theta d\phi^2)
\end{equation}
where \begin{equation}\label{gamma}
  \gamma = \frac{2M}{b},~~~b=2\sqrt{M^2+q^2}
 \end{equation}
are two different parameters which fix the line element. $q$ is
usually so called the scalar charge and $M$ is  mass of the black
hole. It is easy to show that position of the black hole horizons
(apparent and event) is $r_{h}=b.$ In fact, this metric is
solution of Einstein equation in presence of a massless scalar
field with charge $q$ such that
\begin{equation}\label{24}R_{\mu\nu}=8\pi\partial_\mu\phi\partial_\nu\phi,~~~\square\phi=0,~~~
\square=g^{-\frac{1}{2}}\partial_\mu\big(g^\frac{1}{2}g^{\mu\nu}\partial_\nu\big),~~~g=|det
g_{\mu\nu}|\end{equation} with particular real scalar field
solution
\begin{equation}\label{25}\phi(r)=\frac{q}{b\sqrt{4\pi}}\ln\bigg(1-\frac{b}{r}\bigg).\end{equation}
We should note that $b<r<\infty$ in which $b$ is called the
curvature singularity too. To study Hawking radiation of the above
black hole in presence of quantum fluctuations of the above scalar
field, we must in first step, generalize the above scalar field to
be complex with two real and imaginary parts. Then, we should
assume that the complex KG scalar field is a moving wave depended
to both of time and spatial coordinates. To do so, we consider
action functional of a generalized KG scalar field such that
\begin{equation}\label{ac} S=\frac{1}{2}\int \sqrt{g}dx^4\bigg(g^{\mu\nu}\partial_\mu \Phi\partial_\nu
\Phi-\mathfrak{m}^2\Phi^2+\frac{\lambda}{2}\Phi^4\bigg)
\end{equation} in which
$\mathfrak{m}$ is mass of the complex KG scalar field $\Phi$ and
dimensionless parameter $\lambda$ is self-interaction (Higgs)
coupling constant. By varying the action functional (\ref{ac})
with respect to the field $\Phi$ we obtain corresponding equation
of motion such that
\begin{equation}\label{Kel}\square\Phi+\mathfrak{m}^2\Phi-\lambda\Phi^3=0.\end{equation}
 If we set
$\mathfrak{m}=0=\lambda$ and $\Phi=\Phi^*,$ the above
 Euler-Lagrange equation reduces
to the form $\square \Phi=0$ which reads to the equation
(\ref{24}) with particular solution (\ref{25}). At the second
step, to study the Hawking radiation of this quantum black hole,
we choose our methodology
 via creation and
annihilation of quantum scalar particles $\hat{\Phi}$ near the
horizon of the black hole. I.e. the case (b) defined in the
previous section. Before to describe more about this method, we
point here that why the method (c) can not give us the black hole
Hawking temperature same as one which is given in the introduction
section for the Schwarzschild black hole. Because, by substituting
the time-time component of the metric solution (\ref{ds}), namely
$g_{tt}=(1-b/r)^\gamma$ into the equation
$T=\frac{1}{4\pi}\frac{dg_{tt}(r)}{dr}$, we obtain $T=\frac{\gamma
b}{4\pi r^2}(1-b/r)^{\gamma-1}$ which on the horizon surface $r=b$
diverges to infinity because $0<\gamma<1.$ Hence, we must choose
two other methods (a) or (b) which we mentioned in the
introduction section. Use of method (b) is done by calculating
Hamiltonian density operator versus the creation and annihilation
operators and the calculation of the Bogolubov coefficients
between ingoing and outgoing modes of the quantum matter waves. In
fact, the last (self-interaction) part of the action (\ref{ac})
makes nonlinear the equations of motion of the fields $\Phi$ and
$\Phi^*.$ Hence, to obtain local orthogonal mode solutions of the
above mentioned wave equations we must set $\lambda=0$ to obtain
free field solutions near and far from the black hole horizon.
Then, by using the perturbation method, we can calculate effects
of $\lambda\Phi^4$  part in the Hamiltonian density to produce the
Hawking quantum particles. Mathematical calculations of this part
made it lengthened the length of the article and so we decided to
separate it as a new
work and it is addressed in our next work.\\
Spherically symmetric property of the line element (\ref{ds}) lets
us to can decompose the KG scalar field for each mode
$(\omega,\ell, m)$ such that
\begin{equation}\label{28} \Phi_{\omega\ell
m}(t,r,\theta,\phi)=e^{i\omega t}\frac{R_{\omega\ell}(r)}{r}
Y_{\ell m}(\theta,\phi)
\end{equation}
where \begin{equation}Y_{\ell
m}(\theta,\varphi)=\sqrt{\frac{(2\ell+1)}{4\pi}\frac{(\ell-m)!}{(\ell+m)!}}P_{\ell
m}(\cos\theta)e^{im\varphi}\end{equation} are well known spherical
harmonics and dimensionless field $R_{\omega\ell}(r)$ is radial
part of scalar waves which should be determined by KG wave
equation (\ref{Kel}). We must keep $R_{\omega\ell}(r)$ as
dimensionless quantity, because, dimension in the KG field $\Phi$
is inverse of length which it is supported by $r^{-1}$ factor in
the decomposition. By substituting (\ref{28}) and line element
(\ref{ds}) and $\lambda=0,$ into the equation (\ref{Kel}), one can
show that the radial part $R_{\omega\ell}(r)$ is satisfied by the
following equation.
\begin{equation}\label{29}\frac{d^2R_{\omega\ell\gamma}
}{dr^2}+\bigg[\bigg(\frac{1+\gamma}{2}\bigg)\frac{b}{r^2}\bigg(1-\frac{b}{r}\bigg)^{-1}
\bigg]\frac{dR_{\omega\ell\gamma}}{dr}+\bigg[\omega^2\bigg(1-\frac{b}{r}
\bigg)^{-2\gamma}\end{equation}$$-\mathfrak{m}^2\bigg(1-\frac{b}{r}\bigg)^{-\gamma}-\bigg(\frac{(1+\gamma)}{2}\frac{b}{r^3}+
\frac{\ell(\ell+1)}{r^2}\bigg)\bigg(1-\frac{b}{r}\bigg)^{-1}\bigg]R_{\omega\ell\gamma}=0$$
which far from the horizon $r>>b$ reads
\begin{equation}\label{dis}\frac{d^2R_k}{dr^2}+k^2R_k\approx0,~~~k_M(\omega)=\pm\sqrt{\omega^2-\mathfrak{m}^2}.\end{equation}
Asymptotic mode solutions of the above equation are
\begin{equation}\label{210}\lim_{r>>b}R^{\pm}_{k_M}(r)\sim e^{\pm i
k_Mr}\end{equation} where the subscript $M$ denotes to
asymptotically flat (Minkowski) region of the spacetime $r>>b.$ By
substituting these solutions into the relation (\ref{28}), we
obtain normalized modes solutions
\begin{equation}\label{217}\Phi^{M\pm}_{\omega\ell m}=\frac{C_M}{\sqrt{4\pi\omega}}\frac{e^{i(\omega t\pm k_Mr)}}{r}
Y_{\ell,m}(\theta,\varphi)
\end{equation}
in which, we use the arbitrary constant $C_M$ to balance dimension
of the mode solution (\ref{217}) with $\Phi.$ The normalization
coefficient $(4\pi\omega)^{-\frac{1}{2}}$ in the above solution is
calculated by using conserved scalar product \cite{Bir},\cite{Par}
\begin{equation}\label{219}(\chi_+,\chi_-)=i\int_{\Sigma_\eta}\{\chi^*_+\partial_\eta\chi_--\chi_-\partial_\eta\chi^*_+\}
dV_\eta\end{equation} where $\Sigma_\eta$ is a space like
hypersurface which enclose the volume $V_\eta$. In fact, they are
now orthonormal states of a local Hilbert space in $M$ region of
the spacetime and so we can decompose the field $\Phi$ in $M$
region versus the ingoing mode solutions (\ref{217}) such that
\begin{equation}\label{PhiM}\Phi=\sum_{\ell m} \int d\omega\{A_{\omega\ell m}
\Phi_{\omega\ell m}^{M+}+A_{\omega\ell m}^{\dag}\Phi_{\omega\ell
m}^{M+*}\}\end{equation}
 or similarly versus the outgoing waves $\Phi_{\omega\ell
m}^{M-}.$ In the above expansion,  $A$ and $A^\dag$  are
 annihilation and creation  operators of particles
defined on the Hilbert space in $M$ region respectively and their
equal-time commutation relations between the field $\hat{\Phi}$
and its canonical momenta operators $\hat{\Pi}_{\Phi}$ read
\begin{align}\label{dagA}&[A_{\omega\ell,m},A_{\omega'\ell^\prime,m^\prime}]=0,~~~[A_{\omega\ell,m}
^\dag,A_{\omega'\ell^\prime,m^\prime}^\dag]=0,\notag\\&
[A_{\omega\ell,m},A_{\omega^\prime\ell^\prime,
m^\prime}^\dag]=\delta_{\ell\ell^\prime}\delta_{mm^\prime}\delta(\omega-\omega^\prime).\end{align}
 We now investigate similar
analysis to obtain orthogonal mode solutions near the horizon
$r\to b$ region so called as $H$ region. To do so, it is useful
to choose tortoise coordinate such that
\begin{equation}\label{tor}dr^*=\bigg(1-\frac{b}{r}\bigg)^{-\gamma}dr\end{equation} for which
the spacetime line element (\ref{ds}) reads
\begin{equation}ds^2=\bigg(1-\frac{b}{r}\bigg)^{\gamma}dudv-\bigg(1-\frac{b}{r}
\bigg)^{1-\gamma}r^2(d\theta^2+\sin^2\theta
d\varphi^2)\end{equation} where $u=t-r^*$ and $v=t+r^*$ are
retarded and advanced null coordinates respectively. Usually, this
particular coordinates system is useful to study behavior of waves
near the horizon of a spherically symmetric curved spacetime.
However, one can use (\ref{tor}) to transform the equation
(\ref{29}) versus $r^*$ such that
\begin{align}\label{213}&\frac{d^2R_{\omega\ell\gamma}}{dr^{*2}}+\frac{(1-\gamma)}{2}\frac{b}{r^2}\bigg(1-\frac{b}{r}\bigg)^{\gamma-1}
\frac{dR_{\omega\ell\gamma}}{dr^*}+\bigg[\omega^2-\mathfrak{m}^2\bigg(1-\frac{b}{r}\bigg)^\gamma\notag\\&-\bigg(\frac{(1
+\gamma)}{2}\frac{b}{r^3}+\frac{\ell(\ell+1)}{r^2}\bigg)
\bigg(1-\frac{b}{r}\bigg)^{2\gamma-1}\bigg]R_{\omega\ell\gamma}=0.\end{align}
Near the horizon $r\to b,$ this equation reduces to the following
simple form for particular (Schwarzschild) case $\gamma=1.$
\begin{equation}\label{214}\frac{d^2R_{\omega}}{dr^{*2}}+\omega^2R_{\omega}\approx0
\end{equation} with asymptotic mode solutions \begin{equation}\label{H}\lim_{r\to b}R^{\pm}_{k_{H}}(r^*)\sim e^{\pm i k_Hr^*},~~
~k_H(\omega)=\omega,~~~\gamma=1\end{equation} where subscript $H$
denotes to region of `near the Horizon` of the spacetime. Hence,
for charged scalar field $q\neq0,$ in which $0<\gamma<1,$
 we assume
\begin{equation}\label{2199}\lim_{r\to b}R^{\pm}_{k_{H}}(r^*)\sim f_{\gamma}(r) e^{\pm i k_Hr^*},\end{equation} to be mode solution of
the equation (\ref{213})  and $f_\gamma(r)$ satisfies the
following equation.
\begin{align}\bigg(1-\frac{b}{r}\bigg)\frac{d^2f_{\gamma}}{dr^2}&+\bigg[\frac{(1-3\gamma)b}{r^2}\pm i\omega\bigg(1-\frac{b}{r}
\bigg)^{1-\gamma}\bigg]\frac{df_{\gamma}}{dr}\notag\\&+\bigg[\pm
i\omega\frac{(1-\gamma)b}{2r^2}\bigg(1-\frac{b}{r}\bigg)^\gamma-\mathfrak{m}^2\bigg(1-\frac{b}{r}\bigg)^{1-\gamma}\notag\\&
-\bigg(\frac{(1+\gamma)b}{2r^3}+\frac{\ell(1+\ell)}{r^2}
\bigg)\bigg(1-\frac{b}{r}\bigg)^{2\gamma}\bigg]f_{\gamma}=0.\end{align}
Near the horizon $r\to b,$ the above equation reads
\begin{equation}(r-b)\frac{d^2f_{\gamma}}{dr^2}+(1-3\gamma)\frac{df_{\gamma}}{dr}\approx0\end{equation}
because of $0<\gamma<1.$ This equation has a general solution such
that
\begin{equation}f_{\gamma}(r)\sim C_1+C_2 \bigg(\frac{r}{b}-1\bigg)^{3\gamma}\end{equation}
 in which the constant coefficients $C_{1,2}$ could be determined by boundary conditions near the horizon.
 The asymptotic solutions (\ref{H}) show that mass parameter of waves $\mathfrak{m}$ is not
a dominant quantity near the horizon and just the frequency
 of the waves $\omega$ and the dimensionless scalar charge $\gamma$ parameter are
 dominant quantities. For $0<\gamma<1,$ one can see that the second term at the horizon position $r=b$
 vanishes and so we keep
 just $C_1=C_H\neq0$ and $C_2=0$ because, dimension of the field $\Phi$ and mode solution (\ref{2199}) should have same dimensions ($A,B$ and
  $A^\dag, B^\dag$
  are dimensionless). By regarding these
 conditions and by substituting these solutions into the relation (\ref{28}), we
obtain mode solutions in Hilbert space defined in the $H$ region
of the spacetime such that
\begin{equation}\label{221}\Phi^{H\pm}_{\omega\ell m}=\frac{C_H}{\sqrt{4\pi\omega}}\frac{e^{i\omega(t\pm r^*)}}{r}
Y_{\ell m}(\theta,\varphi)
\end{equation}
in which, the normalization coefficient
$(4\pi\omega)^{-\frac{1}{2}}$ is obtained by using scalar product
(\ref{219}) for two dimensional mode solutions (\ref{2199}) on the
hypersurface $u=constant.$ In fact, they are now orthogonal states
of a local Hilbert space in $H$ region of the spacetime. Hence, we
can decompose the field $\Phi$ in $H$ region versus the ingoing
mode solutions $\Phi_{\omega\ell m}^{H+}$ (or similar expansion
for outgoing modes solutions $\Phi_{\omega\ell m}^{H-}$) of the
local Hilbert space in H space such that
\begin{equation}\label{PhiH}\Phi=\sum_{\ell,m} \int d\omega\{B_{\omega\ell m}\Phi_{\omega\ell m}^{H+}+B_{\omega\ell m}^{\dag}
\Phi_{\omega\ell m}^{H+*}\}.\end{equation}
 $B$ and $B^\dag$ in the above expansion are annihilation and
creation operators of particles respectively defined in the
Hilbert space of $H$ region. Their equal-time commutation
relations between the field $\hat{\Phi}$ and its canonical momenta
operators $\hat{\Pi}_{\Phi}$ read
\begin{equation}\label{dagB}[B_{\omega\ell,m},B_{\omega'\ell^\prime m^\prime}]=0,~~~[B_{\omega\ell m}^\dag,B_{\omega'\ell^\prime m^\prime}^\dag]=0
,~~~[B_{\omega\ell m},B_{\omega'\ell^\prime
m^\prime}^\dag]=\delta_{\ell\ell^\prime}\delta_{mm^\prime}\delta(\omega-\omega^\prime).\end{equation}
 Now, we investigate distribution function of created particles
  namely, the number density matrix for the created particles to obtain
   the Hawking temperature of the quantum perturbed JNW black hole in absence of the self-interaction potential $\lambda \Phi^4$.
 \section{Quantum particle creation and  temperature of the JNW black hole}
  In fact, to study creation of particles in curved spacetimes, it is not necessary to solve the equation (\ref{213}) in detail,
  but, our more attention is to observations from communication of asymptotic solutions at far from the horizon ($M$ region) and
  ones which are near the
  horizon ($H$ region).
   Because, the incoming waves $\Phi_{\omega\ell m}^{M,H+}$ will scatter partially back off the gravitational field to become
   a superposition of incoming waves and outgoing waves $\Phi_{\omega\ell m}^{M,H-}.$
   It is obvious that orthonormal states of local Hilbert space in $H$ region are not orthogonal to states of local Hilbert space in $M$ region.
   Hence, they are not a complete set to make a global Hilbert space.
   This is because of curvature of spacetime with no Poincare
symmetry groups same as one which is present in the flat Minkowski
spacetime. In fact, each state of local Hilbert space in $M$
region can be expand versus all states of Hilbert space in $H$
region and vice versa via the Bogolubov transformations such that
\begin{equation}\label{224}\Phi_{\omega}^{M-}=\int d\omega^\prime\{\alpha_{\omega\omega^\prime}\Phi_{\omega^\prime}^{H+}
+\beta_{\omega\omega^\prime}\Phi_{\omega^\prime}^{*H+}\}.\end{equation}
 The Bogolubov transformation relates the operators $A,A^\dag$ to $B,B^\dag$ and vice versa too such that
 \begin{equation}\label{Ob}B_\omega=\int d\omega^\prime\{\alpha^*_{\omega\omega^\prime}A_{\omega'}-\beta^*_{\omega\omega^\prime}
A^\dag_{\omega^\prime}\}\end{equation} and
\begin{equation}\label{O}A_\omega=\int
d\omega^\prime\{\alpha_{\omega\omega^\prime}B_{\omega'}+\beta_{\omega\omega^\prime}
B^\dag_{\omega^\prime}\}\end{equation} where we drop the indexes
$\ell$ and $m$ for simplicity. By substituting the above
expansions, the commutation relations (\ref{dagA}) and
(\ref{dagB}) read to the following identities.
\begin{equation}\label{Pomega}\int
d\omega^\prime(\alpha_{\omega_1\omega^\prime}\alpha^*_{\omega_2\omega^\prime}-\beta_{\omega_1\omega^\prime}\beta^*_{\omega_2
\omega^\prime})=\delta(\omega_1-\omega_2)\end{equation} as
probability conservation equation and
\begin{equation}\int d\omega^\prime(\alpha_{\omega_1\omega^\prime}\beta_{\omega_2\omega^\prime}-\beta_{\omega_1\omega^\prime}
\alpha_{\omega_2\omega^\prime})=0.\end{equation} However, the
coefficients of Bogolubov transformations are determined by the
following scalar products \cite{Bir}, \cite{Par}.
\begin{equation}\label{235}\alpha_{\omega\omega^\prime}=(\Phi^{M-}_{\omega},\Phi^{H+}_{\omega^\prime}),~~~\beta_{\omega\omega^\prime}=
-(\Phi^{M-*}_{\omega},\Phi^{H+}_{\omega^\prime}).\end{equation} We
are now in position to calculate the above Bogolubov coefficients
to obtain rate of created particles near the JNW black hole
horizon which can be observed by detectors located at far from the
horizon. To do so, we are interested in that particular linear
combination of incoming modes $\Phi^{H+}_{\ell m}$ given by
(\ref{221}) that corresponds to standard modes at future null
infinity $J^+$ as $\Phi^{M-}_{\ell m}$. By tracing these modes in
time through the collapsing sphere and out along the advanced null
ray to past null infinity $J^-$, the mode which has the form
$\Phi^{M-}_{\ell m}$ on $J^+$ looks like the following form in
null coordinates systems $\{u,v,\theta,\varphi\}.$
\begin{equation}\label{psiM}\Phi_{\ell m}^{M-}\leftrightharpoons
 \Phi^{H+}_{\ell m (back)}=\frac{C_H}{\sqrt{4\pi\omega}}\frac{e^{-i\omega u(v_H-v)}}{r}Y_{\ell m}(\theta,
\varphi),~~~v<v_H.\end{equation}  In the above wave packet, $v_ H$
is the particular (advance) time where last created particle near
the horizon can be scape to future null infinity $J^+$ before than
the horizon of a collapsing ball is formed. By according the
mathematical calculations in Appendix I, we infer that
\begin{equation}\label{uv}u(v_H-v)\simeq-2b\gamma\ln\bigg(\frac{v_H-v}{\epsilon}\bigg).\end{equation}
For the times $v\geq v_H$, the all advance particle waves are
\begin{equation}\label{psiH}\Phi^{H+}_{\omega\ell m}=\frac{C_H}{\sqrt{4\pi\omega}}\frac{e^{-i\omega v}}{r}Y_{\ell m}(\theta,\varphi),~~~v\geq v_H\end{equation}
which collide with center of collapsing ball and then reflect
back, to reach to the future null infinity $J^+$ but, they can not
reach to $J^+$ regretfully and will have trapped because of
formation of the horizon for the collapsing body. At last, they
are locked up by the newly born black hole. In other words, to
calculate the Bogolubov coefficients (\ref{235}), we should use
the backward moving wave (\ref{psiM}), instead of the outgoing
wave $\Phi_{\ell m}^{M-}$ given by (\ref{217}) and we calculate
(\ref{219}) such that
\begin{align}\alpha_{\omega\omega^\prime}&=(\Phi_{\omega(back)}^{H+},\Phi_{\omega^\prime}^{H+})=i\int_{t=0}(\Phi_{
\omega(back)}^{H+^*}\partial_{r^*}
\Phi_{\omega^\prime}^{H+}-\Phi_{\omega^{\prime}}^{H+}\partial_{r^*}\Phi_{\omega(back)}^{H_-*})dV_{r^*}
\end{align}which by substituting (\ref{psiM}), (\ref{uv}), (\ref{psiH}) and $dV_{r^*}=r^{*2}dr^*\sin\theta d\theta d\varphi$
at far from the black hole horizon $r>>b$ in which $r\approx
r^*\equiv v|_{t=0},$ we obtain
\begin{align}\alpha_{\omega\omega^\prime}&=\frac{C_H^2}{4\pi}\bigg[\bigg(\frac{\omega}{\omega^\prime}\bigg)^\frac{1}{2}
+\bigg(\frac{\omega^\prime}{\omega}\bigg)^\frac{1}{2}\bigg]
\int_{-\infty}^{v_H}dv\bigg(\frac{v_H-v}{\epsilon}\bigg)^{-2b\gamma\omega
i}e^{-i\omega^\prime v}.\end{align} In the above equation, we use
orthogonality condition on the spherical harmonic functions
$Y_{\ell m}(\theta,\varphi),$ such that
\begin{equation}\int_0^{2\pi}d\varphi\int_0^\pi \sin\theta d\theta Y_{\ell m}(\theta,\varphi)Y^{*}_{\ell m}(\theta,\varphi)=1
.\end{equation} By substituting $\xi=(v_H-v)/\epsilon,$ the above
integral equation reads
\begin{align}\label{3133}\alpha_{\omega\omega^\prime}&=\frac{\epsilon C_H^2}{4\pi} e^{-i\omega^\prime v_H}\bigg[
\bigg(\frac{\omega}{\omega^\prime}\bigg)^\frac{1}{2}
+\bigg(\frac{\omega^\prime}{\omega}\bigg)^\frac{1}{2}\bigg]
\int_0^{\infty}d\xi \xi^{-2b\gamma\omega
i}e^{i\omega^\prime\epsilon \xi}\end{align} and by substituting
$i\omega^\prime \epsilon \xi=-\tau,$ the equation (\ref{3133}) can
be written versus the Gamma function:
\begin{equation}\label{Gamma}\Gamma(z-1)=z!=\int_0^\infty \tau^z e^{-z}d\tau,\end{equation}
such that:
\begin{align}\label{alphaa}\alpha_{\omega\omega^\prime}&=
\frac{\epsilon C_H^2}{4\pi}e^{-i\omega^\prime
v_H}\bigg[\bigg(\frac{\omega}{\omega^\prime}\bigg)^\frac{1}{2}
+\bigg(\frac{\omega^\prime}{\omega}\bigg)^\frac{1}{2}\bigg]
\bigg(\frac{-1}{i\epsilon\omega^{\prime}}\bigg)^{1-2b\gamma\omega
i}(-2b\gamma\omega i)!.\end{align}
 By
using the identity (see chapter 10 in Ref. \cite{Arf})
\begin{equation}\label{iden}(is)!(-is)!=|(is)!|^2=\frac{\pi s}{\sinh\pi s},\end{equation}
one can show that (\ref{alphaa}) reads
\begin{equation}\label{3177}\alpha_{\omega\omega^\prime}\alpha^*_{\omega\omega^\prime}=\bigg(\frac{C_H^2}{4\pi}\bigg)^2\bigg[
\bigg(\frac{\omega}{\omega^\prime}\bigg)^\frac{1}{2}
+\bigg(\frac{\omega^\prime}{\omega}\bigg)^\frac{1}{2}\bigg]^2\bigg(\frac{e^{2\pi
b\gamma\omega}}{\omega^{\prime2}}\bigg)\bigg(\frac{2\pi
b\gamma\omega}{\sinh(2\pi b\gamma\omega)}\bigg).\end{equation}
 Similar calculations give us
\begin{align}\label{betaa}\beta_{\omega\omega^\prime}&=-(\Phi_{\omega(back)}^{H+*},\Phi_{\omega^\prime}^{H+})=-i\int_{t=0}
(\Phi_{\omega(back)}^{H+}
\partial_{r^*}\Phi_{\omega^\prime}^{H+}-\Phi_{\omega^\prime}^{H+}\partial_{r^*}\Phi_{\omega(back)}^{H+})
dV_{r_*}\notag\\&=\frac{-
C_H^2\delta_{0m}}{4\pi}\bigg[\bigg(\frac{\omega^\prime}{\omega}\bigg)^\frac{1}{2}+
\bigg(\frac{\omega}{\omega^\prime}\bigg)^\frac{1}{2}
\bigg]\int_{-\infty}^{v_H}dve^{-i\omega^\prime
v}\bigg(\frac{v_H-v}{\epsilon}\bigg)^{2b\gamma\omega i}\end{align}
where we used
\begin{equation}\int_0^{2\pi}d\varphi\int_0^\pi \sin\theta d\theta
Y_{\ell m}(\theta,\varphi)Y_{\ell
m}(\theta,\varphi)=\delta_{0m}.\end{equation} We substitute again
$\xi=(v_H-v)/\epsilon,$  $i\omega^\prime \epsilon\xi=\tau$ and the
identity (\ref{Gamma}), into the integral equation (\ref{betaa})
to obtain
\begin{equation}\label{320}\beta_{\omega\omega^\prime}=\frac{-C_H^2\delta_{0m}\epsilon e^{-i\omega^\prime v_H}}{4\pi}
\bigg[\bigg(\frac{\omega}{\omega^\prime}\bigg)^\frac{1}{2}
+\bigg(\frac{\omega^\prime}{\omega}\bigg)^\frac{1}{2}\bigg]
\bigg(\frac{-1}{i\epsilon\omega^{\prime}}\bigg)^{1+2b\gamma\omega
i}(2b\gamma\omega i)!.\end{equation} By multiplying the above
equation to its complex conjugate and by  substituting the
identity (\ref{iden}), we obtain:
\begin{equation}\label{321}
\beta_{\omega\omega^\prime}\beta^*_{\omega\omega^\prime}=\bigg(\frac{C_H^2}{4\pi}\bigg)^2\bigg[\bigg(\frac{\omega}{\omega
^\prime}\bigg)^\frac{1}{2}
+\bigg(\frac{\omega^\prime}{\omega}\bigg)^\frac{1}{2}\bigg]^2\frac{e^{-2\pi
b\gamma\omega}}{\omega^{\prime2}}\bigg(\frac{2\pi
b\gamma\omega}{\sinh (2\pi b\gamma\omega)}\bigg)\end{equation}
where we substitute $m=0.$ By comparing (\ref{3177}) and
(\ref{321}) we find
\begin{equation}\label{identi}\alpha_{\omega\omega^\prime}\alpha^*_{\omega\omega^\prime}=e^{4\pi
b\gamma\omega}\beta_{\omega\omega^\prime}\beta^*_{\omega\omega^\prime}
\end{equation} which is valid for part of the wave packet that was propagated back in times through the
collapsing body, just before it formed a black hole. Thus, we can
decompose all infalling wave packets
 $\Phi_\omega^{H+}$ to two components as \begin{equation}
\Phi_{\omega^\prime}^{H+}=\Phi_{\omega^\prime}^{H+(1)}+\Phi_{\omega^\prime}^{H+(2)}\end{equation}
where  $\Phi_{\omega^\prime}^{H+(1)}$ and
$\Phi_{\omega^\prime}^{H+(2)}$ are part of wave packets which
propagate disjoint regions on past null infinity $J^-$ for times
$v>v_H$ and $v<v_H$ respectively and thus they are orthogonal. In
other words, $\Phi_{\omega^\prime}^{H+(1)}$ scape out of the
horizon after forming the horizon of a collapsing body to form a
black hole but, $\Phi_{\omega^\prime}^{H+(2)}$ is part of waves
packet which trapped by the horizon and can not to eject to future
null infinity $J^+.$ The latter part can not be observed by an
observer far from the horizon. For this part we can define
`absorptivity` parameter $F(\omega_2)$ such that
\begin{equation}F(\omega_2)\delta(\omega_1-\omega_2)=(\Phi_{\omega_1}^{H+(2)},\Phi_{\omega_2}^{
H+(2)})=\int_0^\infty
d\omega^\prime(\alpha_{\omega_1\omega^\prime}\alpha^*_{\omega_2\omega^\prime}-\beta_{\omega_1\omega^\prime}\beta^*_{\omega_2
\omega^\prime})
\end{equation}
which  in the case $\omega_2=\omega_1=\omega$ and by using the
Fourier transformation
\begin{equation}\delta(\omega_1-\omega_2)=\lim_{\mathcal{T}\to\infty}\frac{1}{2\pi}\int_{
-\frac{\mathcal{T}}{2}}^{\frac{\mathcal{T}}{2}}dte^{it(\omega_1-\omega_2)}
\end{equation}
reads
\begin{equation}\lim_{\mathcal{T}\to\infty}F(\omega)\bigg(\frac{\mathcal{T}}{2\pi}\bigg)=(e^{4\pi
b\gamma\omega}-1)\int_0^\infty
d\omega'\beta_{\omega\omega^\prime}\beta^*_{\omega\omega^\prime}\end{equation}
where we substitute (\ref{identi}). In this view, the total number
of created particles per unit frequency that reach to future null
infinity $J^+$ at late times $\mathcal{T}\to\infty$ in the ejected
wave packet $\Phi_{\omega}^{H+(2)}$ from horizon the JNW black
hole and received by an observer far from the horizon, is
\begin{equation}\label{N}<N_{\omega}>=<0,H|A^\dag_{\omega}A_{\omega}|H,0>=\int_0^\infty
 d\omega^{\prime}|\beta_{\omega\omega^\prime}|^2=\lim_{\mathcal{T}\to\infty}
\bigg[ \frac{(\mathcal{T}/2\pi)F(\omega)}{e^{4\pi
b\gamma\omega}-1}\bigg]
\end{equation}
 and
the absorptivity parameter is
\begin{align}\label{F}F(\omega)&=\int_0^\infty d\omega^\prime(\alpha_{\omega\omega^\prime}\alpha^*_{\omega\omega^\prime}
-\beta_{\omega\omega^\prime}\beta^*_{\omega\omega^\prime}
)\notag\\&=\frac{C_H^4\gamma}{16\pi}\int_0^\infty
d\omega^\prime\frac{(\omega+\omega^\prime)^2}{\omega^{\prime3}}.\end{align}
The eq. (\ref{N}) confirms that the JNW black hole behaves as
`gray` body at temperature $T=\frac{1}{4\pi b\gamma}$ with
absorptivity $F(\omega).$ If $F(\omega)=1,$ then, there is not any
backscattered waves (see \cite{Par} page 169) and so  the JNW
black hole will behave as black body radiation with the Plack`s
distribution (see figure 2-d and compare with figures 2-a,2-b and
2-c). The effect of backscattering is to deplete the outgoing flux
of particles by a factor $1-F(\omega)$ representing reflection
back down the black hole. When we straightforwardly try to
evaluate (\ref{N}), then we encounter with an infinity because of
infinite value for $\mathcal{T}.$ Thus, it is appropriate we
consider the total number created particles per unit frequency and
per unit time as
\begin{equation}\label{n}n_{\omega}=\frac{d<N>}{d\mathcal{T}}=\frac{F(\omega)/2\pi}{e^{4\pi
b\gamma\omega}-1}\end{equation} but, by looking at the equation
(\ref{F}), one can infer that the absorptivity has infinite value
too. In fact, the infinities are because of neglecting the mass
loss of the black hole at duration of particle creation. In other
words, the total number of created particles is infinite because
there is a steady flux of particles reaching to $J^+$ at late
times $\mathcal{T}\to\infty.$ To resolve the infinity of
absorptivity factor in the equation (\ref{F}), we use two ultra
violet cutoff
 $\omega_u$ and infrared cutoff $\omega_i$ for up and down limits of the integration. To do so, we investigate two different approaches
 as follows.
 \subsection{Cutoff frequencies vs particle and black hole masses }
  By regarding mass loss of the evaporating JNW black hole at duration of emission of created particles to $J^+,$
  instead of the equation (\ref{F}),
  we consider
  \begin{align}\label{F1}F(\omega)&=\frac{C_H^4\gamma}{16\pi}\int_{\omega_i}^{\omega_u}
d\omega^\prime\frac{(\omega+\omega^\prime)^2}{\omega^{\prime3}}\end{align}
which by according to the equivalence principal of mass-energy
$`E=\mathfrak{m}c^2=\hbar\omega_i`$ and $Mc^2=\hbar\omega_u$ (in
units $c=\hbar=1$), we can set
\begin{equation}\omega_{i}=\mathfrak{m},~~~~\omega_u=M-\mathfrak{m}\approx M,~~~M>>>\mathfrak{m}.\end{equation} Also, we assume that for
each created particle with positive energy which is emitted to far
from the black hole, there is emitted an anti-particle with
negative energy synchronously which falls inside the black hole
and cause to decrease the black hole mass as $M-\mathfrak{m}.$
This is in fact physical interpretation of quantum vacuum state
fluctuation of the KG quantum scalar field near the horizon which
creates the pair particles-anti particles.  By regarding this and
defining
\begin{equation}\label{Nt}N_t=\frac{M}{\mathfrak{m}}\end{equation} which
 can
be interpreted as `total internal particles that make up a JNW
black hole with mass $M$` we find
\begin{align}F(\omega)&\approx\frac{C_H^4\gamma}{16\pi}\bigg[\ln N_t+\frac{2\omega}{\mathfrak{m}}+
\frac{1}{2}\bigg(\frac{\omega}{\mathfrak{m}}\bigg)^2\bigg]
\end{align}
 for $N_t>>1$ but, accessible. It is important to note that $F(\omega)$ is depend to
 physical content of the black hole, "i.e.
 the type of particles
 inside the black hole" which make up
it but, the distribution function $(e^{4\pi b\gamma
\omega}-1)^{-1}$ given in the equation (\ref{n}), is free of
structure of the particles and the black hole. Denominator factor
in the distribution function (\ref{N}) shows that the JNW black
hole has temperature
\begin{equation}\label{TT}T=\frac{1}{4\pi b\gamma}=\frac{1}{8\pi\gamma M_{JNW}}\end{equation} in units $\hbar=c=G=K_B=1$, in which
$K_B$ is Boltzmann`s constant. By looking at (\ref{TT}), one can
infer that temperature of a JNW black hole is larger than the
Schwarzschild black hole temperature such that
\begin{equation}\frac{T_{JNW}}{T_{Sch}}=\frac{1}{\gamma}=\frac{\sqrt{M^2+q^2}}{M}>1,~~~0<\gamma<1\end{equation}
where we assumed that the mass of JNW black hole is equal to a
compared Schwarzschild black hole mass $M=M_{JNW}=M_{Sch}.$ Total
luminosity of each black hole such as the JNW quantum perturbed
black hole is given by the following expression (see \cite{Par}
and references therein).
\begin{align}L&=\frac{1}{2\pi}\sum_{\ell=0}^\infty (2\ell+1)\int_0^\infty n_{\omega}|\tau_{\ell\omega}|^2\omega  d\omega\notag\\&
=\bigg(\frac{C_H^4\gamma}{16\pi}\bigg)\bigg(\frac{1}{2\pi}\bigg)^2\bigg(\sum_{\ell=0}^\infty
(2\ell+1)|\tau_{\ell\omega}|^2\bigg)\notag\\&\times\frac{1}{(8\pi
M)^2}\bigg[\ln
N_t\int_0^\infty\frac{xdx}{e^x-1}+\frac{2}{\mu}\int_0^\infty\frac{x^2dx}{e^x-1}+\frac{1}{2\mu^2}\int_0^\infty\frac{x^3dx}{e^x-1}\bigg]
\end{align} in which \begin{equation}\label{def}x=4\pi b\gamma \omega,~~~4\pi b\gamma \mathfrak{m}=\mu=8\pi M\mathfrak{m}\end{equation}
 and $\tau_{\ell\omega}$ is transmission
coefficient for a wave of unit amplitude incident from infinity .
It can be computed by numerical integration of the eq. (\ref{29})
(see \cite{Par}). We can show easily
\begin{equation}\label{mu}\mu\approx2\pi\gamma\end{equation} because,
 each wave packet of created particles with mass $\mathfrak{m},$ is
trapped by the new born black hole and so its de Brogle wave
length
$\lambda=\frac{\hbar}{\mathfrak{m}c}=\frac{1}{\mathfrak{m}}$ (in
units $\hbar=c=1$,) should be in order of the black hole horizon
radius $\lambda\approx 2b.$ Hence, by substituting
 these equalities into the
definition (\ref{def}), one find
 (\ref{mu}).
In the above equation, integral parts can be calculated via
`residue calculation` as
\begin{equation}\int_0^\infty\frac{x^{s}dx}{e^x-1}=s!\zeta(s+1)\end{equation} where $\zeta(s)$ is Riemann Zeta
function (see ref. \cite{Arf})
\begin{equation}\label{zet}\zeta(s)=\sum_{n=1}^{\infty}\frac{1}{n^s}, ~~~s>1
.\end{equation}In summary, we see that total luminosity of the
evaporating JNW black hole become
\begin{equation}\label{LM}L=\frac{C_L}{M^2}\end{equation}
in which we set
\begin{align}C_L&
=\bigg(\frac{C_H^4\gamma}{128}\bigg)\bigg(\frac{1}{2\pi}\bigg)^5\bigg(\sum_{\ell=0}^\infty
(2\ell+1)|\tau_{\ell\omega}|^2\bigg)\notag\\&\times\bigg[\zeta(2)\ln
N_t+\frac{\zeta(3)}{\pi\gamma}+\frac{\zeta(4)}{8\pi^2\gamma^2}\bigg]
\end{align}
 with $\zeta(2)\approx1.65,~\zeta(3)\approx1.2$ and
$\zeta(4)\approx1.1$ (see table 5.3 in ref. \cite{Arf}). The
coefficient $C_L$ is calculated by Page (1976) for neutrinos,
photons, and gravitons as $16.36\times10^{-5},$
$3.37\times10^{-5}$ and $0.38\times10^{-5}$ respectively (see
\cite{Par} and references therein). To study effects of total
number of particles content of the black hole $N_t$ and $\gamma$
factor of the metric on the luminosity, we define a dimensionless
luminosity per unit frequency and per unit time as
\begin{equation}\label{dL}\frac{dL^*}{dx}=\frac{x\ln N_t+x^2/\pi\gamma+x^3/8\pi^2\gamma^2}{e^x-1}\end{equation} where
 \begin{equation}L^*=\frac{L}{\big(\frac{C_H^4\gamma}{16\pi}\big)\big(\frac{1}{2\pi}\big)^2\big(\sum_{\ell=0}^\infty
(2\ell+1)|\tau_{\ell\omega}|^2\big)}.\end{equation} We plotted
diagram of the equation (\ref{dL}) by  figures 1 and 2 for
different
 values of the parameters $\gamma$ and $N_t.$
\subsection{Cutoff frequencies vs birth time of black hole $v_H$}
In this section, we re-calculate (\ref{F}) again but, by different
method as follows. We know that $v_H$ is the time for which the
last created particle (the wave packet) can be scape to future
null infinity $J^+$ before than to appear the new born JNW black
hole from a collapsing body. Hence, this trapped waves form some
standing waves inside the black hole and so we can make discrete
all modes $\omega^\prime$ instead of continues form in the
equations (\ref{alphaa}) and (\ref{320}). To do so, we keep
$e^{-i\omega^\prime v_H}=e^{-ik\pi}$ with $k=1,2,3,\cdots.$
 In this case we find
 \begin{equation}\omega^\prime_k=\frac{k\pi}{v_H}\end{equation}
 for which we can replace the integrations in (\ref{F1}) with summations
 such that
 \begin{align}F(\omega)&=\frac{C_H^4\gamma}{16\pi}\bigg[\omega^2\int_{\omega_i}^{\omega_u} \frac{d\omega^\prime}{\omega^{\prime3}}+2\omega\int
 _{\omega_i}^{\omega_u} \frac{d\omega^\prime}{\omega^{\prime^2}}+\int_{\omega_i}^{\omega_u} \frac{d\omega^\prime}{\omega^\prime}
  \bigg]\notag\\&\approx
\frac{C_H^4\gamma}{16\pi}\bigg[\bigg(\frac{v_H\omega}{\pi}\bigg)^2\sum_{k=1}^{\infty}\frac{1}{k^3}
+\frac{2\omega v_H}{\pi}\sum_{k=1}^{\infty}\frac{1}{k^2}
+\sum_{k=1}^{\infty}\frac{1}{k}\bigg]\notag\\&=\frac{C_H^4\gamma}{16\pi}\bigg[\bigg(\frac{v_H\omega}{\pi}\bigg)^2\zeta(3)
+\frac{2\omega v_H}{\pi}\zeta(2)+\zeta(1)\bigg]\
\end{align}
in which $\zeta(x)$ is the zeta function (\ref{zet}) and
$\zeta(1)$ called as Harmonic series is indeterminate (or
non-convergent) (see chapter 5 in ref.\cite{Arf}). This property
of non-convergency caused us to abandon this method and use the
previous method given in subsection (3.1) to use graphical studies
of the work in figures 1 and 2. Because, at least $N_t$ gives the
meaning of the number of particles that make up the black hole,
which is more acceptable.  In the following, we calculate mass
loss equation of the quantum unstable JNW black hole.
\section{Mass loss for quantum  JNW black hole}
The energy conservation principle leads us to use (\ref{LM}) for
mass loss equation of the evaporating quantum perturbed JNW black
hole such that
\begin{equation}-\frac{dM}{dt}=\frac{C_L}{M^2},~~~C_L>0\end{equation} which
means that decrease of a black hole mass (the negative sign in
left side of the above equation) in time is balance with
luminosity or emission of Hawking radiation of created particles.
This equation is valid till before than the mass of the
evaporating JNW black hole reaches to the Planck mass for which
quantum fields in curved space become invalid and after than the
full quantum gravity is dominant. At this regime, one can obtain a
suitable classical solution for the above mass loss equation as
\begin{equation}\frac{M(t)}{M_0}=\bigg(1-\frac{t}{t_{\infty}}\bigg)^\frac{1}{3}\end{equation}
in which
\begin{equation}t_{\infty}=\frac{M_0^3}{3C_L}\end{equation}
is total time for which all initial mass of the black hole $M_0$
disappear because of Hawking evaporation.
\section{Interpretation of diagrams}
We plotted mass loss of the quantum perturbed JNW black hole in
figure 1-a. This shows decrease of mass of the black hole because
of its luminescence and its total mass disappears completely at a
finite time. In fact, we know from Heisenberg uncertainty
principal which this black hole can not lost its total mass
because, at final state of evaporation, the backreaction stress
tensor of the created particles make stop the loss of the mass
such that final remnant mini black hole should be remain as a
stable (see for instance \cite{Ghaf1}). By looking at the
relations (\ref{def}) and (\ref{mu}), one can obtain an
uncertainty relation between the black hole mass and the KG
quantum particle mass such that
\begin{equation}\label{un}\mathfrak{m}M=\frac{\gamma}{4}\end{equation}
which right side comes from the scalar charge of the classical JNW
black hole. This relationship prevents the black hole from losing
all its mass. By eliminating the KG particle mass $\mathfrak{m}$
between (\ref{un}) and (\ref{Nt}), one find the Bekenstein-Hawking
entropy of the JNW black hole which is equivalent with square of
the black hole mass. To do so, we have
\begin{equation}N_t=\frac{4}{\gamma}M^2=\gamma b^2=\frac{\gamma}{\pi}\bigg(\frac{\mathcal{A}}{4}\bigg).\end{equation}
where $\mathcal{A}$ is surface area of the JNW black hole horizon
and so we interpret $\pi N_t/\gamma$ to be Bekenstein-Hawking
entropy of the
JNW black hole. \\
Figure 1-b shows luminosity per unit frequency and per unit time
at constant $\gamma=\frac{1}{30}$ for different population of
particles that make up black hole $N_t$. By raising $N_t,$ one can
see that pick of the diagram takes higher values. By comparing the
figures 1-b with 1-c and 1-d, one can infer that increase of
$\gamma$ factor (which we interpret dimensionless scalar charge of
the JNW black hole) causes that position of maximum value (the
pick point) of the diagram
 decreases. One can see that shape of these diagrams differ versus to diagram of a Planck`s black body radiation energy density
 distribution given by figure 2-d.  This is because of shape of absorptivity factor $F(\omega)$ which causes
  the quantum perturbed
 JNW black hole to be have a gray body radiation. This means that some parts of
 spectrums are absorbed by the black hole and
 do not emitted never far from the horizon to future null infinity $J^+,$ while, a black body cavity emit all of spectrums which absorbed.
 By looking at the figures 2-a and 2-b and 2-c, one find that pick
 of the luminosity per unit frequency of the JNW black hole dose
 not changed by raising total particles content of the black
 hole.\\
 After to discovery of thermal
radiation of the black holes by the  Hawking \cite{Haw1},
\cite{Haw2}, the region of origin of the radiated quanta has been
a topic of debate. In summary, there have been proposed two
different scenario to answer to this question presented.  First of
them is "firewall" argument which means that  the conjectured lack
of maximal UV-dependent entanglement between the Hawking pairs
which makes the near horizon state singular and eventually demands
some drastic modification of the near horizon geometry \cite{Alm}
(see also \cite{Ghaf2} in which the UV cutoff frequency has been
set by the cosmological parameter). This is the popular belief
that these Hawking quantum pairs originate from the ultra high
energy excitations very close to the horizon but, the second
scenario propose that their origin is "quantum atmosphere" of
black hole namely a region well outside the  black hole horizon.
Also, the first proposal remains from the times of black hole
Hawking radiation discovery but it seems that the second proposal
is near to fact.\\ The latter scenario argued by Giddings when he
calculated the effective radius of a radiating body via the
Stefan-Boltzmann law \cite{Gidd} and showed that this effective
radius is approximately 79 times to the horizon radius. Next,
further explore of the Giddings issue investigated in the paper
\cite{Dey} to end up corroborating the claim of the Giddings, by
using both a heuristic argument and a detailed study of the stress
energy tensor. In this work authors showed that maximum point of
luminosity or energy density of radiation is happened at radius 2
times more than the Schwarzschild radius which is other reason to
obey the claim of the Giddings.\\ By according to the second
scenario, one can see diagrams of the figure 2 in the present work
where peak point of the luminosity is formed at points $x=2$ and
$x=3$ for $\gamma=\frac{1}{20}$ and $\gamma=\frac{1}{30}$
respectively. The diagrams  do not show a suitable peak for
$\gamma=\frac{1}{10}.$ In other words, smaller values for $\gamma$
make larger dimensionless radius where the peak point appears for
the luminosity diagram. Furthermore, one can see diagram in figure
1-c such that, the position of maximum luminosity moves to larger
values by raising the number of created particles per unit
frequency and per unit time $N_t.$ In summary, our work satisfies
predictions of the previous works \cite{Gidd} and \cite{Dey} such
that the created Hawking pairs particles position is not vicinity
of the black hole horizon but, is several times of the horizon
position and it is depended to smaller values of the dimensionless
charge of the scalar KG quantum wave $\gamma\leq\frac{1}{10}.$
\section{Concluding
remark} In this work, we considered a quantum massive KG scalar
field minimally interacting with JNW spherically symmetric static
black hole. Then, we investigated the Hawking temperature of pair
created particles near the horizon which emitted far from the
horizon. The specific form of the metric prevented us from
obtaining the temperature of the quantum JNW black hole by
calculating the surface gravity. Hence, we calculate Bogolubov
transformation components to find number density function or total
number of created particles per unit frequency and per unit time.
We obtained an absorptivity factor which shows that this black
hole has not black body radiation behavior but it is a gray body
which emits
 part of created particles to future null infinity $J^+.$ While,
other part of the created particles can not scape to infinity and
they are trapped by a new born JNW black hole made from collapsing
body. We calculated temperature of this evaporating quantum JNW
black hole which is hotter than a Schwarzschild black hole with
same ADM mass. This is because of presence of scalar field charge.
At last, we calculated total luminosity of the black hole which is
related to the black hole mass as inverse square function. By
considering energy conservation, we investigated backreaction of
these created quantum particles to give a time-dependent equation
for mass loss of the emitting black hole. It is found that the
black hole lost its whole mass at a finite time. Total time for
disappearances of the black hole is depend to initial mass of the
black hole and this result is valid just before than the decrease
of the black hole mass reduces to the Planck scale. Because, for
times less than the Planck`s time $10^{-34}s,$ the approach of
"quantum field theory in curved space" is invalid and the unknown
pure quantum gravity should be used instead of. This is shown here
by deriving a duality condition between the black hole mass and
the KG particle mass so called as the "uncertainty relation". It
should be noted that results of this work are extracted by
considering the free quantum KG fields which are interacting with
a JNW black hole minimally. As an extension of this work, we will
calculate $\Lambda \Phi^4$ self interaction counterpart of these
free KG fields in our subsequent work and we will investigate
physical effects of this interaction part in rate of particles
creation and possibly of change of the
  quantum perturbed JNW black hole temperature.
\section{Appendix I}
To study moving of radial rays (particles) in approximation of
geometric optics, it is useful to set $\theta=\frac{\pi}{2}$ in
the metric equation (\ref{ds}) because of its spherical symmetry
property. Its timelike killing vector field gives us
\begin{equation}\label{E}E=\bigg(1-\frac{b}{r}\bigg)^{\gamma}\frac{dt}{d\lambda}\end{equation}
and its spacelike killing vector field reads
\begin{equation}\label{L}P_{\varphi}=-\bigg(1-\frac{b}{r}\bigg)^{1-\gamma}r^2\frac{d\varphi}{d\lambda}.\end{equation}
In the above relations $\lambda$ is affine parameter and the
constant energy $E=P_t$ is time-component of four-momentum of
particles and the constant angular momentum $P_\varphi$ is
azimuthal-spatial component of four-momentum of the particle. They
are obtained from the Lagrangian of the particles such that
\begin{equation}\mathfrak{L}_{particles}=\frac{ds^2}{d\lambda^2}=\bigg(1-\frac{b}{r}\bigg)^\gamma \dot{t}^2 -\bigg(1-\frac{b}{r}\bigg)^{-\gamma}
\dot{r}^2
 -\bigg(1-\frac{b}{r}\bigg)^{1-\gamma} r^2 \dot{\phi}^2\end{equation}
derived  from the line element (\ref{ds}) by substituting
$\theta=\frac{\pi}{2}.$ In this lagrangian, we defined
four-velocity components of the scalar particles as
\begin{equation}V^t=\dot{t}=\frac{dt}{d\lambda},~~~V^r=\dot{r}=\frac{dr}{d\lambda},~~~V^\varphi=\dot{\varphi}=\frac{d\varphi}{d\lambda}.
\end{equation}
For light rays with vanishing rest mass, we know $ds=0$ and so
$\mathfrak{L}_{particles}=0$ but, for massive particles with
nonzero rest mass $E_0=m_0c^2\neq0,$ we have
$\mathfrak{L}_{particles}=E^2_0$ in the geometric units $G=c=1.$
However, by substituting the conditions (\ref{E}) and (\ref{L})
into the above Lagrangian, we obtain
\begin{equation}E^2=\dot{r}^2+E_0^2\bigg(1-\frac{b}{r}\bigg)^\gamma+\frac{L^2}{r^2}\bigg(1-\frac{b}{r}\bigg)^{2\gamma-1}\end{equation}
which for radial moving particles we have $L=0$ and so it can be
rewritten as follows.
\begin{equation}\label{36}\frac{dr}{d\lambda}=\pm E\sqrt{1-\frac{E_0^2}{E^2}\bigg(1-\frac{b}{r}\bigg)^\gamma}
\end{equation}
where $+$ is for outgoing and $-$ is for ingoing particles
respectively. Also, by substituting (\ref{E}) and $L=0$ into the
above Lagrangian, we can write
\begin{equation}\label{37}\frac{du}{d\lambda}=\frac{dt}{d\lambda}-\frac{dr^*}{d\lambda}=E\bigg(1-\frac{b}{r}
\bigg)^{-\gamma}\bigg[1\mp\sqrt{1-\frac{E_0^2}{E^2}\bigg(1-\frac{b}{r}\bigg)^{\gamma}}\bigg].\end{equation}
By comparing the asymptotic radial solutions (\ref{210}) and
(\ref{H}), one can infer that the rest mass of the scalar
particles $\mathfrak{m}$ is  not dominated near the horizon
because it is disappeared in $k_H.$ This is seen at the geometric
approximation, by calculating limit of the equations (\ref{36})
and (\ref{37}) near the horizons $r\to b,$ such that
\begin{equation}\label{38}\frac{dr}{d\lambda}\approx\pm E\end{equation}
and
\begin{equation}\label{399}\frac{du}{d\lambda}\approx2E\bigg(1-\frac{b}{r}\bigg)^{-\gamma}\end{equation}
for ingoing particles $-$ and
\begin{equation}\frac{du}{d\lambda}\approx\frac{1}{2}\frac{E_0^2}{E}\end{equation}
for outgoing particles $+$ respectively. To evaluate the Hawking
particles at future null infinity coming from ingoing particles
which they were scape from past null infinity (center of
collapsing ball before to make the horizon) we must consider
(\ref{399}) to evaluate the quantity $u(v_0-v).$ To do so, we
integrate (\ref{38}) for ingoing particles $-$ such that
\begin{equation}r-b\approx-E\lambda\end{equation}
where we assume initial condition $r(0)=b$, namely, the affine
parameter takes zero value on the horizon. This equation gives us
\begin{equation}\bigg(1-\frac{b}{r}\bigg)^{-\gamma}=\bigg(1-\frac{b}{E\lambda}\bigg)^\gamma\end{equation}
which by substituting into the equation (\ref{399}) we obtain
\begin{equation}\label{313}u(\lambda)=2b\int\bigg(1-\frac{1}{\xi}\bigg)^\gamma d\xi,~~~\xi=\frac{E\lambda}{b}.\end{equation}
To obtain a suitable relation for $u(v)$ to be independent of
affine parameter $\lambda,$ we calculate again
\begin{equation}\frac{dv}{d\lambda}=\frac{dt}{d\lambda}+\frac{dr^*}{d\lambda}=E\bigg(1-\frac{b}{r}\bigg)^{-\gamma}
\bigg[1-\sqrt{1-\frac{E_0^2}{E^2}\bigg(1-\frac{b}{r}
\bigg)^\gamma}\bigg]\end{equation} for incoming particles which
near the horizon $r\to b$ reduces to the following simple form.
\begin{equation}\frac{dv}{d\lambda}\approx\frac{E_0^2}{2E}.\end{equation}
By integrating the above equation, we have
\begin{equation}\label{316}v-v_H=\frac{E_0^2}{2E}\lambda=\frac{b}{2}\frac{E_0^2}{E^2}\xi,~~~v<v_H\end{equation}
where we set constant of integration as $v=v_H$ at $\lambda=0,$
corresponding  to $r=b$ for position of the horizon. In this
solution, $v_H$ is advanced particular time for last particle
which can scape from the collapsing ball to future null infinity
and make the created Hawking particles far from the black hole. To
calculate the integral equation (\ref{313}), we can now substitute
the relation (\ref{316}) into the equation (\ref{313}) such that
\begin{equation}\label{317}u(v-v_H)=\frac{2b}{\epsilon}\int\bigg(1+\frac{\epsilon}{v_H-v}\bigg)^\gamma dv\end{equation}
where we defined
\begin{equation}\epsilon=\frac{b}{2}\bigg(\frac{E_0}{E}\bigg)^2.\end{equation}
This integral equation has not an analytic solution and so we
solve it by perturbation series method. By looking at the relation
(\ref{gamma}) we know
that\begin{equation}\gamma=\frac{M}{\sqrt{M^2+q^2}}\end{equation}
and so $0<\gamma<1.$ Also for a relativistic particles, we know
$E_0<E$ and so we use first order Taylor series expansion versus
$\gamma$ and $\epsilon$ to calculate the above integral equation
such that
\begin{equation}\bigg(1+\frac{\epsilon}{v_H-v}\bigg)^\gamma\approx1+\frac{\epsilon\gamma}{(v_H-v)}\end{equation}
 for which (\ref{317})
reads
\begin{equation}u(v_H-v)\simeq-2b\bigg[\frac{(v_H-v)}{\epsilon}+\gamma\ln\bigg(\frac{v_H-v}{\epsilon}\bigg)\bigg]\end{equation}
 which its dominant terms at
$v\to v_H$ reduces to the following form.
\begin{equation}u(v_H-v)\simeq-2b\gamma\ln\bigg(\frac{v_H-v}{\epsilon}\bigg).\end{equation}Apparently, it seems
 that the equation
(\ref{316}) is invalid for light rays with zero rest mass $E_0=0,$
but in page of 158 of the reference \cite{Par}, author presented
an important argument when he calculated same calculations given
in this appendix but, for light rays with no rest mass moving on
Schwarzschild background. In his calculations he showed that there
is general constants instead of the coefficient $\frac{E_0^2}{2E}$
which can be fixed with initial condition of the system.

\begin{figure}
\begin{center}
\subfigure[{}]{\label{39}
\includegraphics[width=0.45\textwidth]{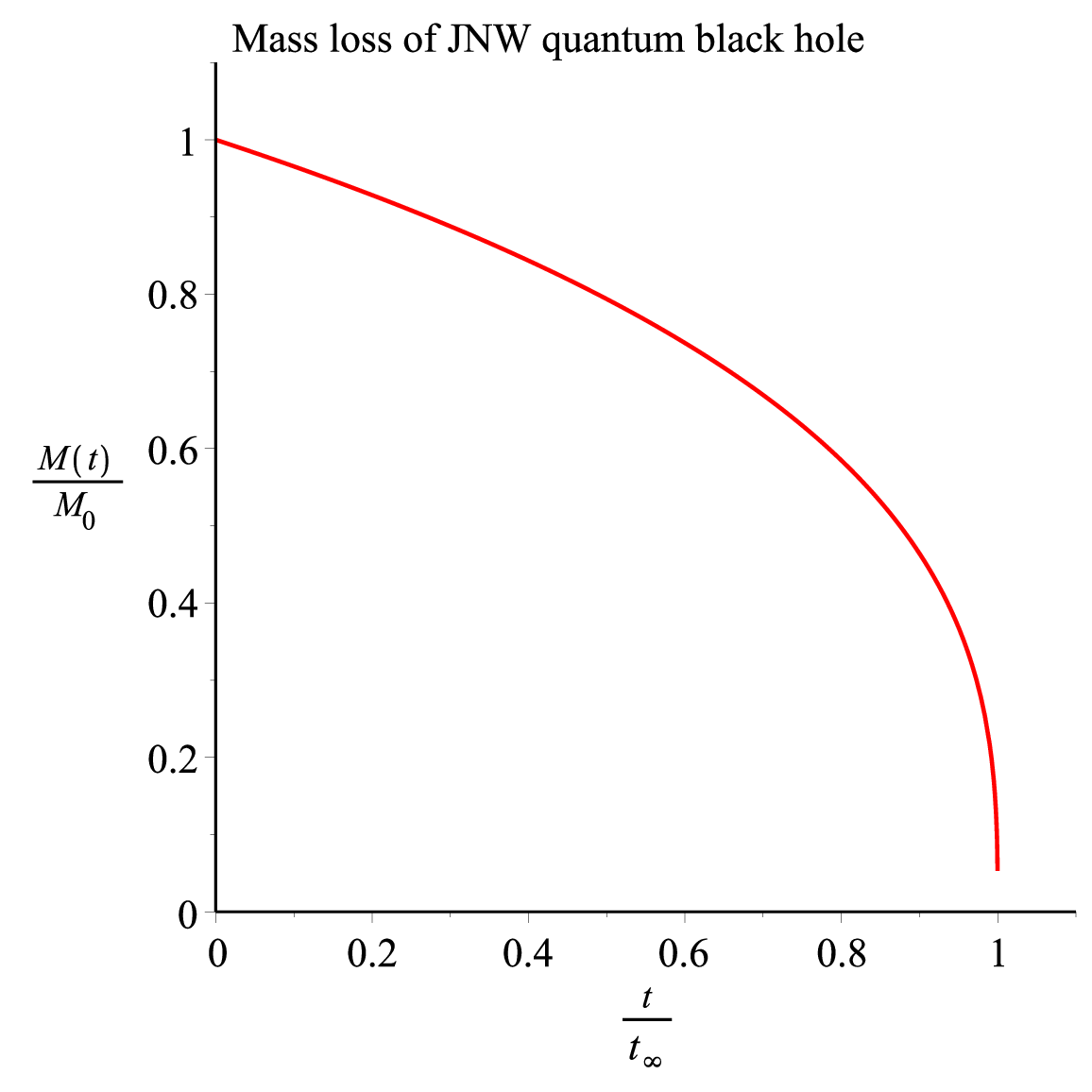}}
\hspace{2mm}\subfigure[{}]{\label{23d413}
\includegraphics[width=0.45\textwidth]{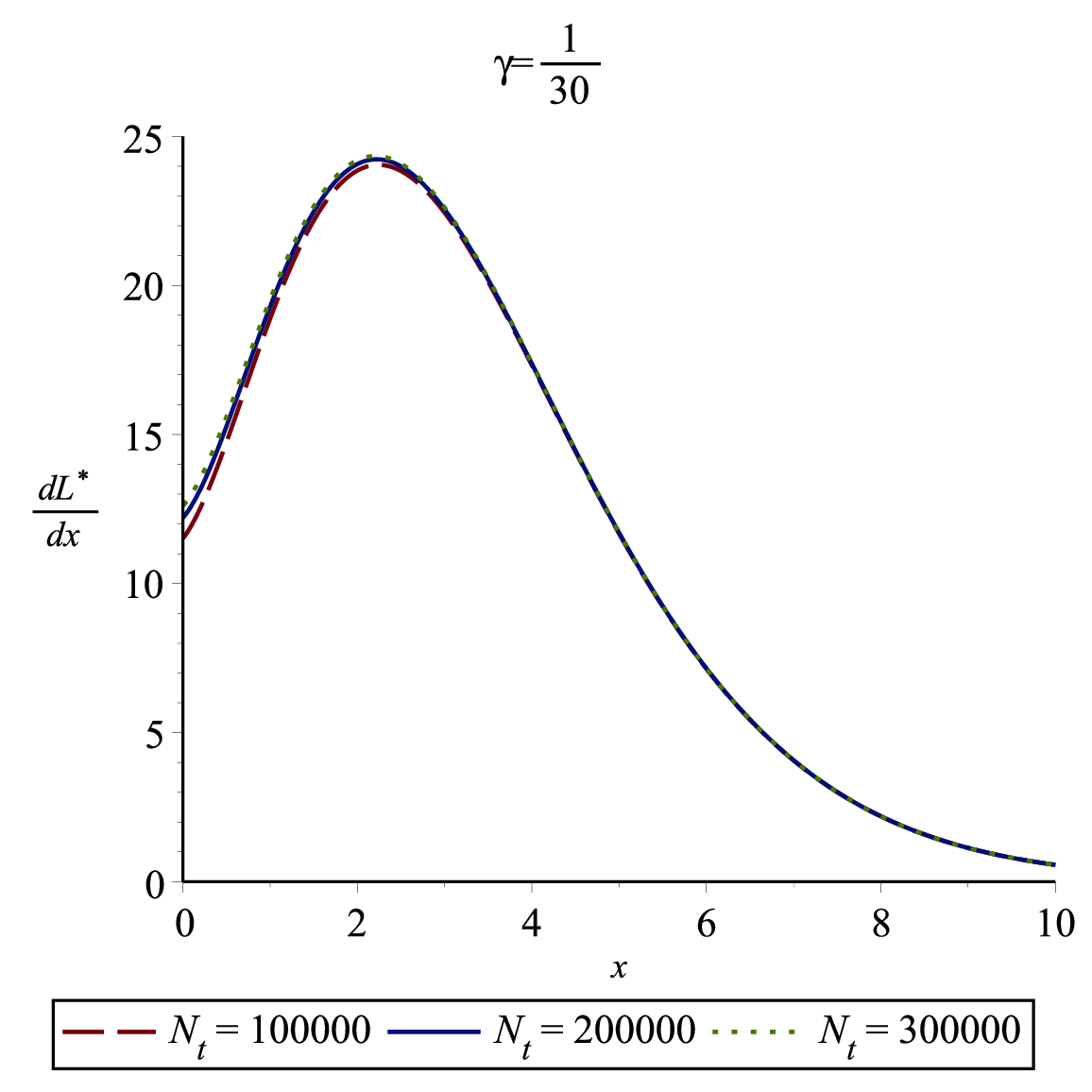}}
\hspace{2mm}\subfigure[{}]{\label{23d36}
\includegraphics[width=0.45\textwidth]{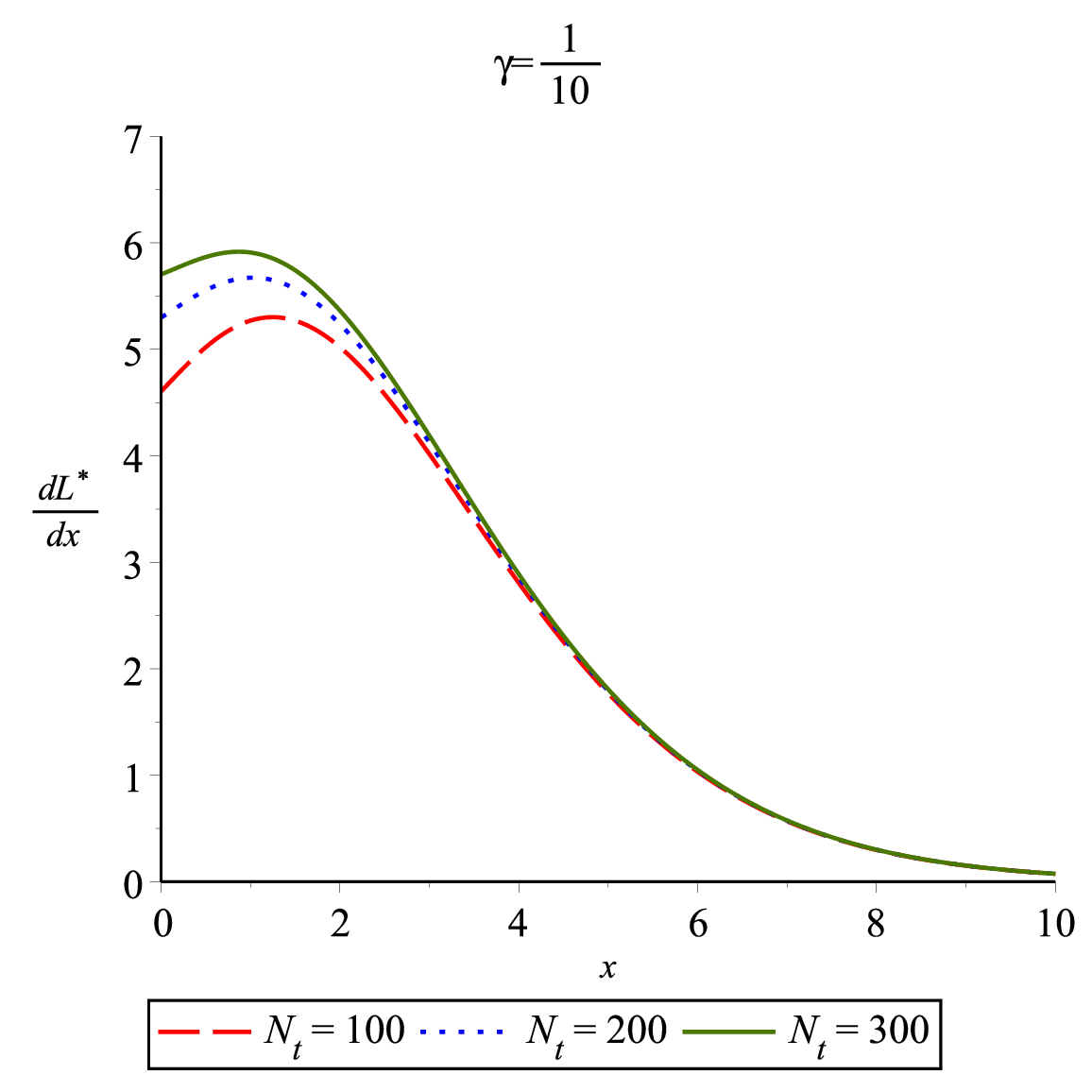}}
\hspace{2mm}\subfigure[{}]{\label{23d336}
\includegraphics[width=0.45\textwidth]{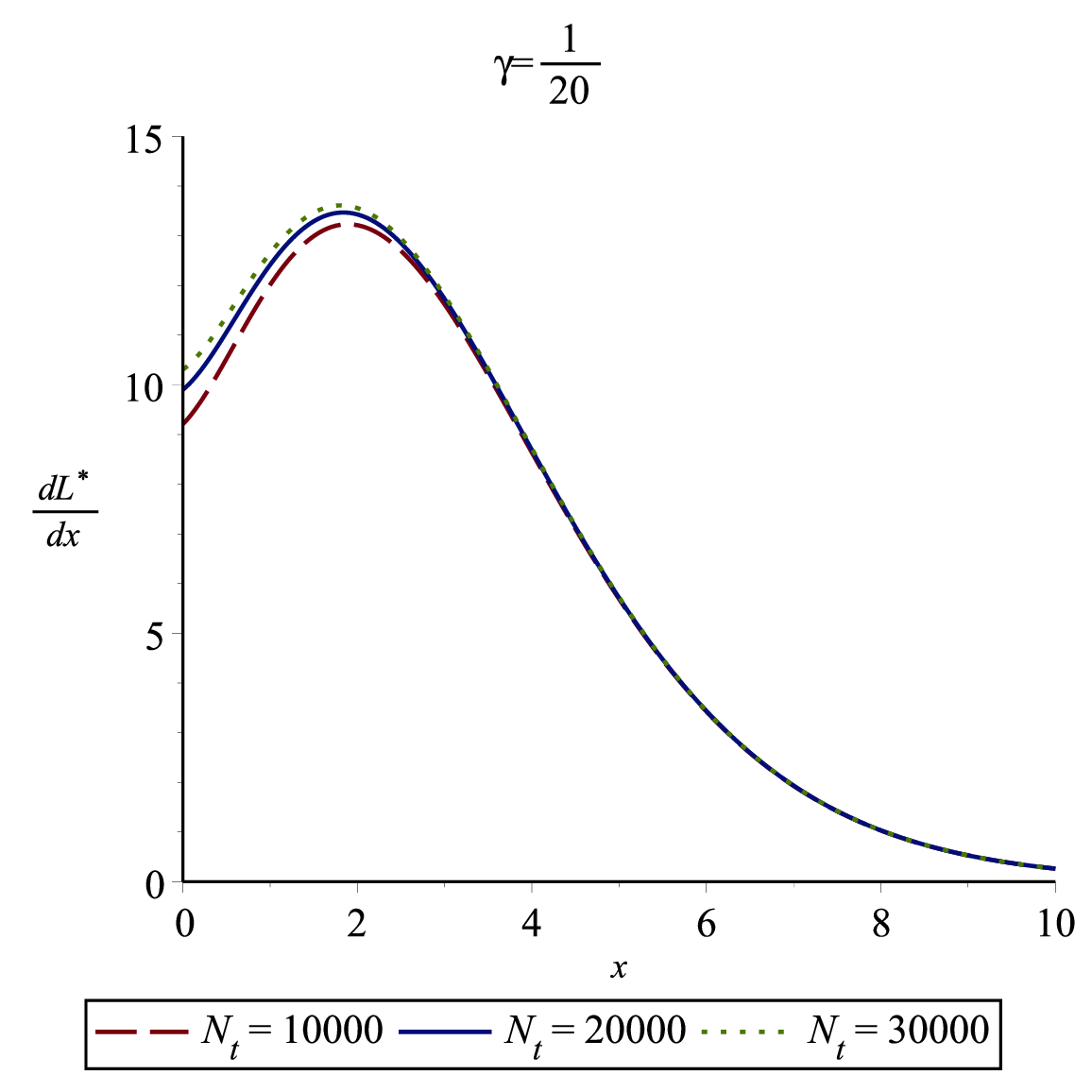}}
\end{center}
\caption{\footnotesize{(a) Mass loss of quantum perturbed JNW
black hole, (b), (c) and (d) JNW black hole luminosity per unit
frequency and per unit time at constant metric parameter $\gamma$
for different $N_t.$}}
\end{figure}
\begin{figure}
\centering\subfigure[{}]{\label{1011}
\includegraphics[width=0.45\textwidth]{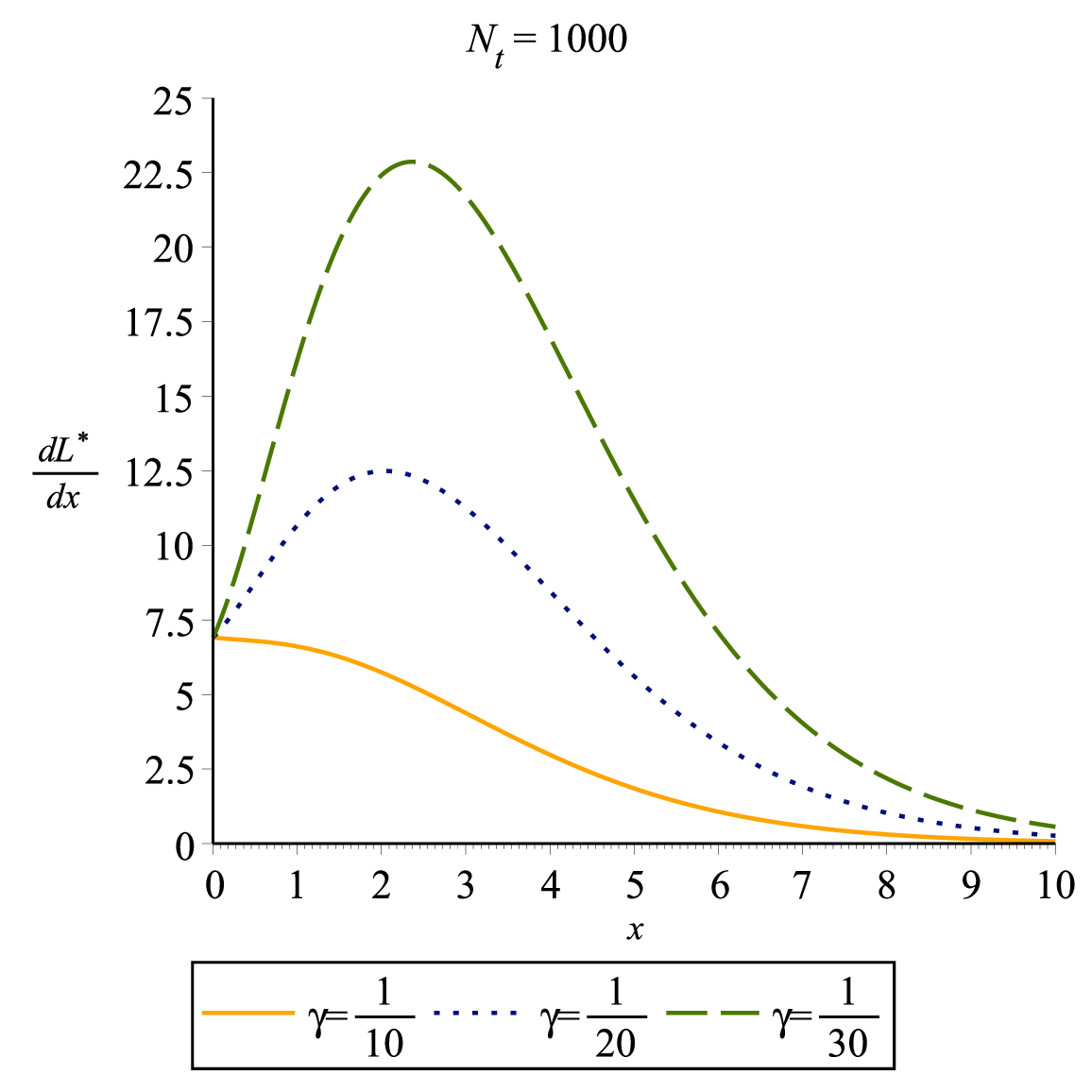}}
\hspace{2mm}\subfigure[{}]{\label{23s411}
\includegraphics[width=0.45\textwidth]{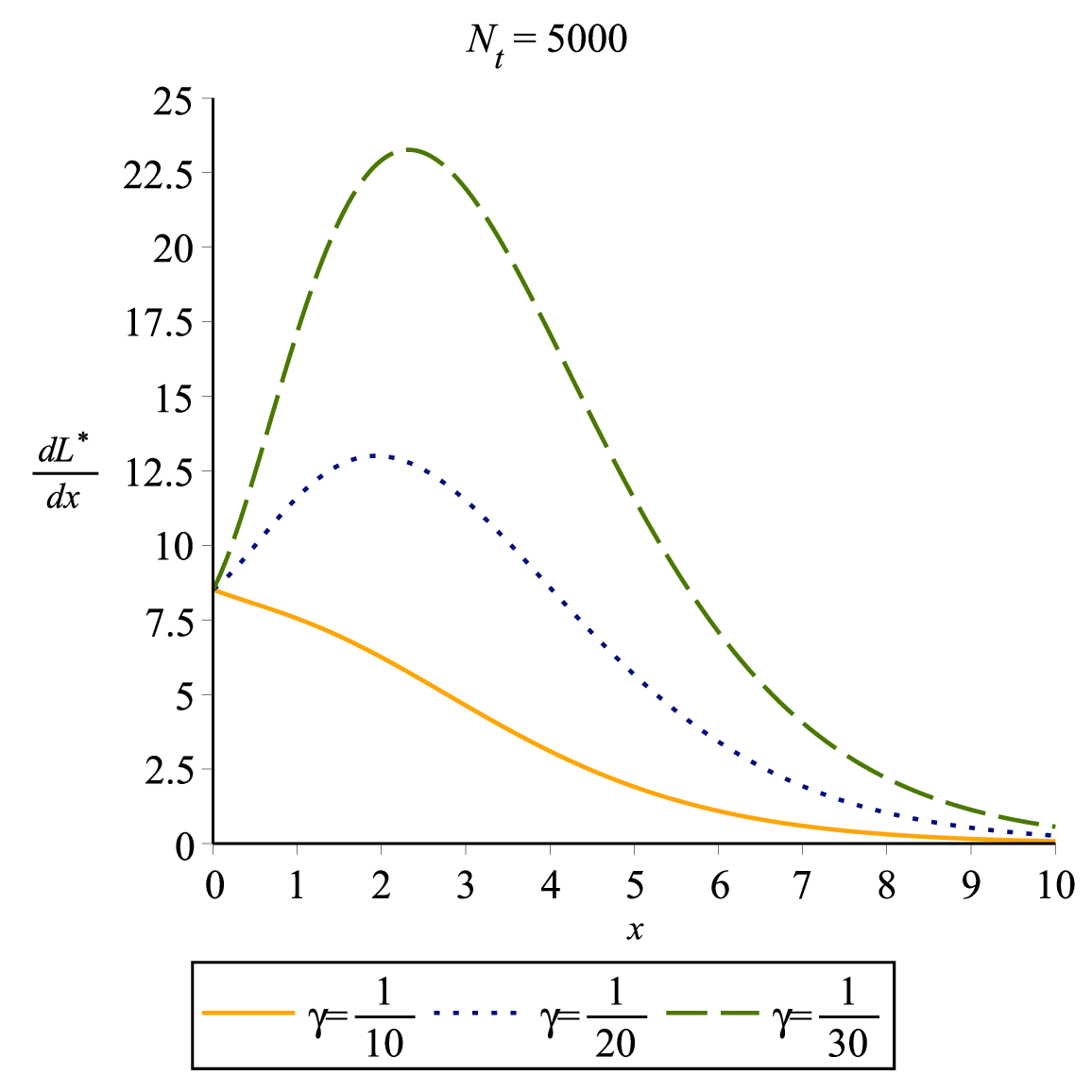}}
\hspace{2mm}\subfigure[{}]{\label{23ss411}
\includegraphics[width=0.45\textwidth]{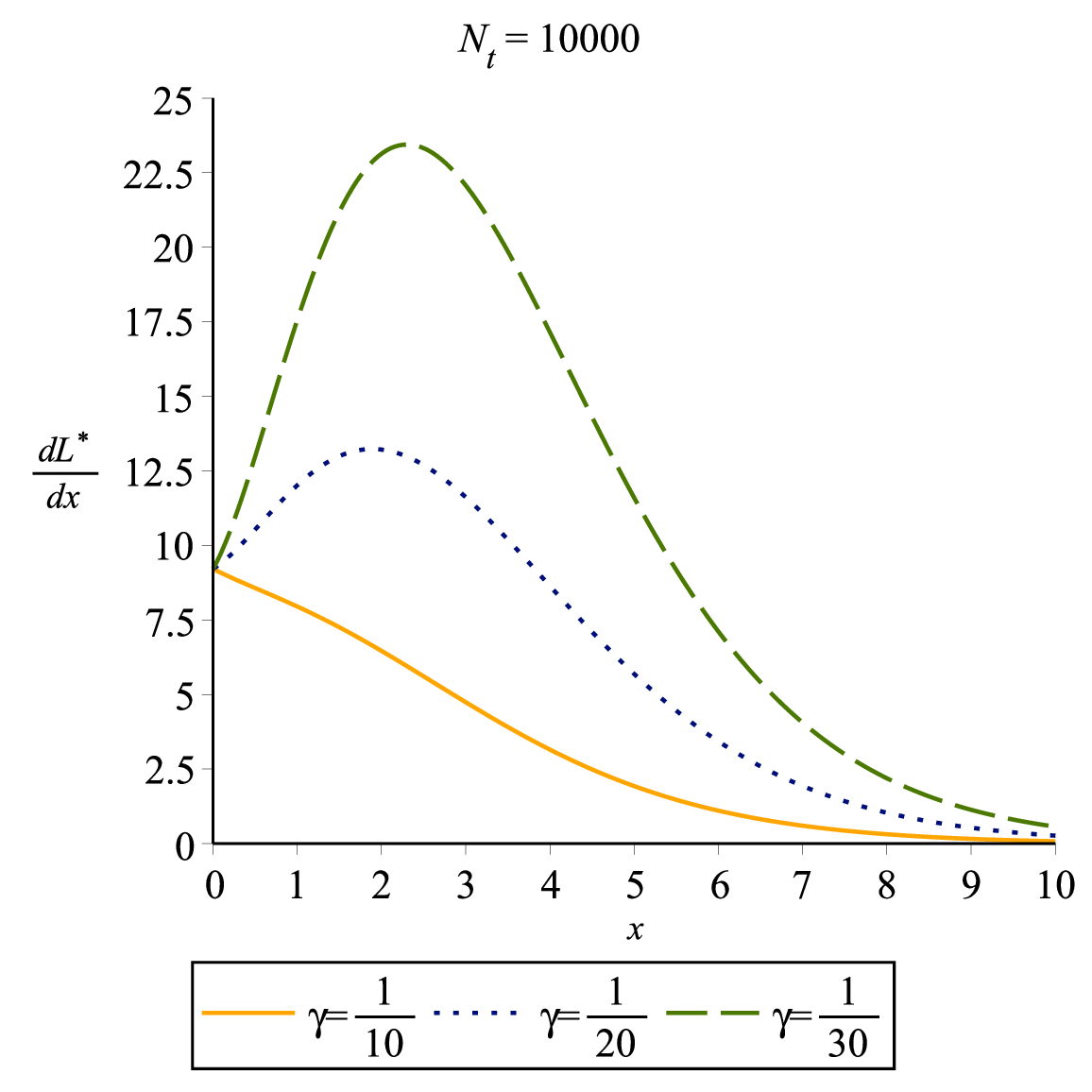}}
\hspace{2mm}\subfigure[{}]{\label{23sss411}
\includegraphics[width=0.45\textwidth]{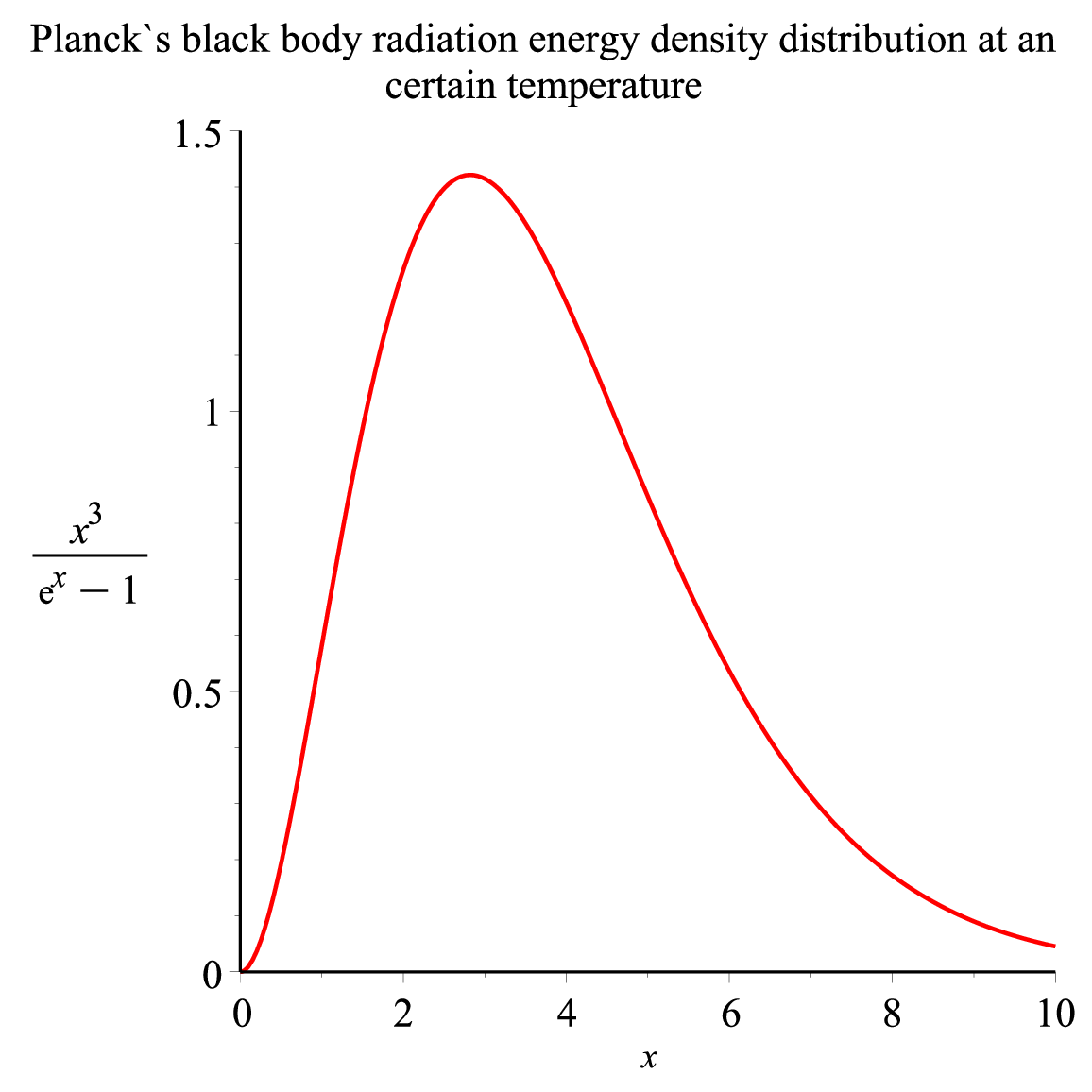}}
\hspace{2mm} \caption{\footnotesize{(a), (b) and (c) Luminosity
per unit frequency and per unit time at constant $N_t$ for
different metric parameter (d) Planck`s black body radiation
energy density distribution at an certain temperature (the
cavity)} (See each quantum physics book for instance figure 11,
chapter 1, ref. \cite{Eis})}
\end{figure}
\end{document}